\documentclass[pre,nopacs,nokeys,nofootinbib,floatfix]{revtex4}
\usepackage{amsfonts}
\usepackage{amsmath}
\usepackage{amssymb}
\usepackage{bbm}
\usepackage[utf8]{inputenc}
\usepackage[caption=false]{subfig}
\usepackage{tensor}
\usepackage{slashed}
\usepackage[centertableaux]{ytableau}
\usepackage{hyperref}
\usepackage{natbib}
\usepackage{braket,mleftright}
\usepackage{graphicx}
\usepackage{cleveref}
\usepackage{tikz}
\usepackage{booktabs}
\usepackage{tikz} 
\usepackage{tikz-feynman}
\usepackage{tikz}
\usepackage{subcaption} 

\begin{document}

\title{Stochastic Field Theory of HIV Latency: \\ Instanton Dynamics and the Path to Viral Rebound}

\author{Jos\'e de Jes\'us Bernal-Alvarado}
\email{bernal@ugto.mx}
\affiliation{Physics Engineering Department, Universidad de Guanajuato, M\'{e}xico}

\author{David Delepine}
\email{delepine@ugto.mx}
\affiliation{Physics Department, Universidad de Guanajuato, M\'{e}xico}

\author{Georges Delepine}
\email{georges.delepine@uclouvain.be}
\affiliation{De Duve Institute, Universit\'e Catholique de Louvain, Belgium}

\date{\today}

\begin{abstract}
The transition from clinical latency to active HIV infection is 
a stochastic process.  Traditional deterministic
models based on ordinary differential equations (ODEs) fail to capture the
extinction boundary and the subsequent rebound of the virus because they
neglect demographic noise in the small-population limit.  We map the resulting Master
Equation onto a coherent-state path integral using the Doi-Peliti
second-quantization formalism and verify that the stochastic Hamiltonian
reproduces the correct mean-field ODEs on the classical manifold
$\bar{\phi}_i = 1$.  In the semiclassical (large system-size) limit we derive
eight coupled Hamilton-Jacobi equations, identify the virus-free and endemic
fixed points, and obtain the basic reproduction number $R_0$ for the
four-species network.  For the analytically tractable case of single-virion
bursting ($n=1$) we show that the stochastic Hamiltonian \emph{factorizes}
into two bilinear terms, exposing a non-trivial zero-energy surface that
constitutes the instanton trajectory.  A quasi-steady-state reduction then
yields a first-order linear ODE whose closed-form solution gives the
instanton action $S_{\rm inst}$ in terms of biological parameters.
Calibrating to clinical data---infected-cell half-life $\delta_I^{-1}\approx
2\,\rm d$, viral burst size $N\approx 10^3$, clearance rate $\delta_V\approx
23\,\rm d^{-1}$, reactivation rate $\eta\approx 10^{-3}\,\rm d^{-1}$, and
latent reservoir size $L_0\approx 10^6$ cells---we obtain a mean first-passage
time (MFPT) to viral rebound of $\approx 6$ months for a typical patient,
with a range of days to years depending on reservoir size.  These predictions
are qualitatively consistent with observed post-interruption rebound
timescales.
\end{abstract}

\maketitle
\section{Introduction}
There is a well-documented clinical paradox associated to HIV infections. Under antiretroviral therapy (ART), HIV viral load decays exponentially to undetectable levels, yet stopping treatment almost always leads to viral rebounds. This suggests that the observed ''eradication'' is not a truly stable state but a metastable one where a latent reservoir of infected cells persists and can spontaneously reactivate. Classical Ordinary differential equation (ODE) based models \cite{Perelson1993} capture the initial exponential decay of viral load but they break down at the extinction boundary. When viral copy numbers are very low ( few latent cells, infected cells and virions), the mean-field approximation loses physical meaning since demographic noise dominates at this moment. And this generates a contradiction: ODEs predict stable eradication, while clinical observations show inevitable rebound after treatment interruption\cite{Perelson1996,Rong2009, Pinkevych2015,Conway2019}.  The origin of this rebound is the
\emph{latent reservoir}: a pool of long-lived, transcriptionally silent
CD4$^+$ T cells that harbor integrated HIV provirus and that sporadically
reactivate~\cite{Finzi1997,Chun1997,Siliciano2003}. Each reactivation event has a nonzero
probability of triggering a self-sustaining infection chain before the immune
system can suppress it.

In stochastic models where the demographic noise dominates, the Doi-Peliti formalism gives us a way to introduce naturally the effect of this noise and to go beyond ODE approximation~\cite{Doi1976,Peliti1985,Kamenev2011}.

Stochastic models of latent HIV have been explored using Gillespie-type
simulations~\cite{Gillespie1977,Rong2009} and branching-process
approximations~\cite{Conway2019,Hill2018,VanDorp2020}, but a rigorous field-theoretic treatment
capable of producing closed-form analytical predictions has remained lacking.
In statistical physics, the problem of escape from a metastable state is
elegantly handled by \emph{instanton calculus}: the path-integral method
originally due to Doi~\cite{Doi1976} and Peliti~\cite{Peliti1985} maps the
Master Equation onto an action functional, whose saddle point---the
instanton---gives the exponentially dominant contribution to the escape rate.

Recently, Taye~\cite{Taye2026} reformulated viral rebound as a
first-passage problem for a Poisson shot-noise process, in which
individual reactivation events contribute bursts to the total plasma
viral load until the cumulative signal crosses an assay detection
threshold $V_{\rm det}$.   The present work is distinct.  First, we derive the stochastic dynamics from a
\emph{microscopic} four-species reaction network via the Doi-Peliti
path integral, so that the basic reproduction number $R_{0}$ and the
rebound probability $P_{\rm rebound}=1-1/R_{0}$ emerge from first
principles. Second, the instanton
framework remains valid and exhibits qualitatively distinct behaviour in
the strong-reactivation regime $\sigma\gtrsim\delta_{I}$ (reservoir
size $L_{0}\gtrsim 500$ cells/mL), where the exponential factor in the
escape rate dominates and mean-field descriptions break down.

The paper is organized as follows.  Section~\ref{sec:model} defines the
four-species stochastic reaction network.  Section~\ref{sec:doipeliti}
constructs the Doi-Peliti Hamiltonian and verifies consistency with the
known mean-field ODEs.  Section~\ref{sec:HJ} derives the Hamilton-Jacobi
equations, identifies the fixed points, and obtains $R_0$.
Section~\ref{sec:instanton} solves the instanton for $n=1$.
Section~\ref{sec:results} calibrates the theory to clinical data and
presents predictions for the MFPT to viral rebound.
Section~\ref{sec:discussion} discusses implications and extensions.

\section{The Four-Species Stochastic Model}
\label{sec:model}
 
We model HIV reservoir dynamics as a Markovian birth-death process involving
four discrete populations: uninfected CD4$^+$ target T cells ($T$), latently
infected cells ($L$), productively infected cells ($I$), and free virions
($V$).  Including $T$ as a dynamic species is essential for two reasons: 
\begin{enumerate}
    \item the consumption of target cells
upon infection must be explicitly tracked for probability conservation in the
Doi-Peliti path integral;
\item CD4$^+$ T-cell depletion is itself a
clinically observable signature of reservoir activity.
\end{enumerate}
Introducing these four populations, to build the effective action describing their interactions, one needs to first detail the complete set of allowed elementary transitions. To construct these interactions, we must rely  on the biological studies of how HIV virus works. Upon contact between a free virion and a CD4$^+$ T cell, the outcome of
infection is not deterministic: it depends critically on the \emph{activation
state} of the target cell at the moment of encounter.  Activated T cells
harbor the transcription factors (NF-$\kappa$B, AP-1, NFAT) required to
drive efficient HIV transcription, and infection of such cells produces a
productively infected cell at rate $\beta(1-f)$.  By contrast, quiescent or
resting T cells lack these factors; reverse transcription stalls after
nuclear import of the pre-integration complex, and the integrated provirus
enters a transcriptionally silent state---the latent reservoir---at rate
$\beta f$.So, the parameter $f\in[0,1]$ represents the latency fraction.

\begin{align}
\varnothing &\xrightarrow{\;\lambda\;} T
  & &\text{(thymic production)} \label{r1}\\
T &\xrightarrow{\;\delta_T\;} \varnothing
  & &\text{(natural T-cell death)} \label{r2}\\
T + V &\xrightarrow{\;\beta f\;} L
  & &\text{(latency-establishing infection)} \label{r3}\\
T + V &\xrightarrow{\;\beta(1-f)\;} I
  & &\text{(productively infecting)} \label{r4}\\
L &\xrightarrow{\;\eta\;} I
  & &\text{(latent reactivation)} \label{r5}\\
L &\xrightarrow{\;\delta_L\;} \varnothing
  & &\text{(latent cell death)} \label{r6}\\
I &\xrightarrow{\;\,k\;\,} I + nV
  & &\text{(viral burst)} \label{r7}\\
I &\xrightarrow{\;\delta_I\;} \varnothing
  & &\text{(infected cell death)} \label{r8}\\
V &\xrightarrow{\;\delta_V\;} \varnothing
  & &\text{(virion clearance)} \label{r9}
\end{align}
The estimated values of the parameters of each reaction are presented in Table (\ref{tab:params}). Each productively infected cell is a viral factory: it transcribes, assembles,
and builds HIV particles at a high rate until it is killed by the immune
system or cytopathic effect.  In the Doi-Peliti reaction (\ref{r7}),
$k$ is the \emph{rate of burst events} and $n$ is the \emph{number of virions
released per event}, so the total virion production rate per infected cell is
$nk\;\mathrm{d}^{-1}$.  Over an infected-cell lifetime $\delta_I^{-1}\approx
2\;\mathrm{d}$, the expected cumulative progeny is the \emph{burst number}
$N = nk/\delta_I \approx 10^3$ virions per cell~\cite{Perelson1996}.
Empirical estimates based on the ratio of latently
to productively infected cells during primary infection place
$f\approx 10^{-3}$--$10^{-5}$; in our central calibration we follow
Rong \& Perelson~\cite{Rong2009} in using $f=10^{-4}$.  The two infection
channels (\ref{r3}) and (\ref{r4}) thus partition the total infection flux
$\beta\phi_T\phi_V$ between the slow reservoir compartment and the fast
productive compartment in the ratio $f:(1-f)$.

Table~\ref{tab:params} summarizes the parameters, their biological
interpretation, and clinical values compiled from the literature.
 
\begin{table}[h]
\caption{Model parameters calibrated to clinical data.}
\label{tab:params}
\begin{tabular}{llll}
\begin{tabular}{llll}
Symbol & Meaning & Central value & Ref. \\
\hline
$\lambda$   & T-cell production rate & $10^4\,\text{mL}^{-1}\text{d}^{-1}$ & \cite{Perelson1996} \\
$\delta_T$  & T-cell death rate & $0.01\,\text{d}^{-1}$ & \cite{Perelson1996} \\
$\beta$     & Infection rate & $2.4\!\times\!10^{-8}\,\text{mL}\,\text{v}^{-1}\text{d}^{-1}$ & \cite{Perelson1996} \\
$f$         & Latency fraction & $10^{-4}$ & \cite{Rong2009} \\
$\delta_L$  & Latent cell death & $4\!\times\!10^{-3}\,\text{d}^{-1}$ & \cite{Pinkevych2015} \\
$\eta$      & Reactivation rate & $10^{-3}\,\text{d}^{-1}$ & \cite{Pinkevych2015} \\
$\delta_I$  & Infected cell death & $0.5\,\text{d}^{-1}$ & \cite{Perelson1996} \\
$k$         & Virion production rate & $500\,\text{d}^{-1}$ & derived$^\dagger$ \\
$\delta_V$  & Viral clearance & $23\,\text{d}^{-1}$ & \cite{Perelson1996} \\
$L_0$       & Latent reservoir & $1\,\text{cell\,mL}^{-1}$ & \cite{Pinkevych2015} \\
\end{tabular}
\end{tabular}
{\small $^\dagger k = N\delta_I$ with burst size $N\approx 10^3$.}
\end{table}
 
\section{The Doi-Peliti Mapping}
\label{sec:doipeliti}
 
\subsection{From Master Equation to Path Integral}
 
Let $P(\bm{n};t)$ be the probability of observing $\bm{n}=(n_T,n_L,n_I,n_V)$
particles at time $t$.  We introduce bosonic creation and annihilation
operators $\hat{a}_i^\dagger$, $\hat{a}_i$ for each species
$i\in\{T,L,I,V\}$, satisfying $[\hat{a}_i,\hat{a}_j^\dagger]=\delta_{ij}$.
The probability state is encoded as
$|P\rangle = \sum_{\bm{n}} P(\bm{n};t)\prod_i (\hat{a}_i^\dagger)^{n_i}|0\rangle$,
and the Master Equation takes the operator form
$\partial_t|P\rangle = \hat{\mathcal{L}}|P\rangle$.
 
Following Doi~\cite{Doi1976} and Peliti~\cite{Peliti1985}, we evaluate the
transition amplitude in the coherent-state basis.  Setting
$\bar{\phi}_i \equiv \hat{a}_i^\dagger$ (the response, or conjugate, field)
and $\phi_i$ (the state field), the evolution of the system is governed by
the action
\begin{equation}
  S[\phi,\bar\phi] = \int dt\!\left[\sum_i \bar\phi_i\,\partial_t\phi_i
   - \mathcal{H}(\phi,\bar\phi)\right].
\label{eq:action}
\end{equation}
 
For each elementary reaction we read off the Doi-Peliti contribution to
$\mathcal{H}$ using the rule: a reaction with stoichiometry
$\bm{\nu}^{\rm in}\!\to\!\bm{\nu}^{\rm out}$ at rate $r$ contributes
$r\!\left(\prod_j \bar\phi_j^{\nu^{\rm out}_j} - \prod_j \bar\phi_j^{\nu^{\rm in}_j}\right)\!\prod_j \phi_j^{\nu^{\rm in}_j}$.
Applying this rule to reactions (\ref{r1})--(\ref{r9}) we obtain the
\emph{normal-ordered stochastic Hamiltonian}

\begin{align}
\mathcal{H} &= \lambda(\bar\phi_T - 1)
 + \delta_T(1-\bar\phi_T)\phi_T \notag\\
&\quad + \beta\!\left[f\bar\phi_L + (1-f)\bar\phi_I - \bar\phi_T\bar\phi_V\right]\phi_T\phi_V \notag\\
&\quad + \eta(\bar\phi_I - \bar\phi_L)\phi_L \notag\\
&\quad + \sum_{X\in\{L,I,V\}}\delta_X(1-\bar\phi_X)\phi_X \notag\\
&\quad + k\,\bar\phi_I\!\left(\bar\phi_V^n - 1\right)\phi_I.
\label{eq:hamiltonian}
\end{align}

The mean-field manifold in this convention corresponds to
$\bar\phi_i = 1$ for all $i$~\cite{Peliti1985}.
The equations of motion $\dot\phi_i = \partial\mathcal{H}/\partial\bar\phi_i$
evaluated at $\bar\phi_i=1$ must reproduce the known ODEs:

\begin{enumerate}
    \item Target cells $T$.
    \begin{equation}
        \dot\phi_T\big|_{\bar\phi=1} = \lambda - \delta_T\phi_T - \beta\phi_T\phi_V
    \end{equation}
    This is the standard T-cell equation of Perelson et al.~\cite{Perelson1996},
with source $\lambda$, natural death $\delta_T$, and infection sink $\beta TV$.
    \item Latent cells $L$.
    \begin{equation}
        \dot\phi_L\big|_{\bar\phi=1} = \beta f\phi_T\phi_V - (\eta+\delta_L)\phi_L
    \end{equation}
    The latent compartment, with influx $\beta f TV$ from the latency-establishing
infection channel and outflux via reactivation ($\eta$) or death ($\delta_L$),
matches Eq.~(1b) of Rong \& Perelson~\cite{Rong2009}.
    \item Infected cells $I$.
    \begin{equation}
      \dot\phi_I\big|_{\bar\phi=1} = \beta(1-f)\phi_T\phi_V + \eta\phi_L
 - \delta_I\phi_I + k\underbrace{(1^n-1)}_{=\,0}\phi_I.  
    \end{equation}
  The burst term \emph{vanishes identically} on the mean-field manifold for
any $n$, correctly encoding that infected cells are neither created nor
destroyed by virion production.  The remaining terms reproduce Eq.~(1c)
of Rong \& Perelson~\cite{Rong2009} and Eq.~(2) of Conway \&
Perelson~\cite{Conway2019}. 
\item Free virions $V$.
\begin{equation}
    \dot\phi_V\big|_{\bar\phi=1} = nk\phi_I - \delta_V\phi_V - \beta\phi_T\phi_V
\end{equation}
Virion production at rate $nk$ per infected cell, clearance at rate $\delta_V$,
and absorption by target cells at rate $\beta TV$ reproduce the standard
viral dynamics equation of Perelson et al.~\cite{Perelson1996} (their
Eq.~(3)), with the additional infection-sink term made explicit.
\end{enumerate} 
The complete four-species ODE system
$\{T,L,I,V\}$ in this form is also studied numerically in
Conway \& Perelson~\cite{Conway2019} and serves as the mean-field backbone
of our stochastic analysis.

\section{Hamilton-Jacobi Equations and Fixed Points}
\label{sec:HJ}

The coherent-state path integral in Eq.~(\ref{eq:action}) is extensive in
the system size $N$: in the limit $N\to\infty$ the integral over all
trajectories is dominated by the single path $(\phi^*,\bar\phi^*)$ that
extremizes the action, with all other trajectories suppressed by
$e^{-cN}$~\cite{AssafMeerson2017,ElgartKamenev2004} where $c$ is constant.  This is the
stochastic analogue of the WKB (semiclassical) approximation in quantum
mechanics, with $1/N$ playing the role of $\hbar$.
 
Setting $\delta S = 0$ with respect to independent variations of $\phi_i$
and $\bar\phi_i$ yields the canonical \emph{Hamilton equations}
\begin{equation}
\dot\phi_i = \frac{\partial\mathcal{H}}{\partial\bar\phi_i},
\qquad
\dot{\bar\phi}_i = -\frac{\partial\mathcal{H}}{\partial\phi_i},
\label{eq:hamilton_canonical}
\end{equation}
one pair for each species $i\in\{T,L,I,V\}$.  Together these eight
first-order ODEs govern the full stochastic dynamics at leading order in
$1/N$.  We seek the \emph{trajectories} $(\phi_i(t),\bar\phi_i(t))$ that
satisfy these equations and connect the biologically relevant fixed
points of the system.  Among all such trajectories, the mean-field
solution ($\bar\phi_i=1$ for all $i$) recovers the deterministic ODEs. 

The semiclassical (large system-size $N$) limit is dominated by the
saddle-point trajectories $\delta S = 0$, yielding the Hamilton-Jacobi
system:
\begin{subequations}
\label{eq:HJ}
\begin{align}
\dot\phi_T &= \lambda - \delta_T\phi_T
  - \beta\bar\phi_V\phi_T\phi_V, \label{HJ:phiT}\\
\dot\phi_L &= \beta f\phi_T\phi_V - (\eta+\delta_L)\phi_L, \label{HJ:phiL}\\
\dot\phi_I &= \beta(1-f)\phi_T\phi_V + \eta\phi_L - \delta_I\phi_I
  + k(\bar\phi_V^n-1)\phi_I, \label{HJ:phiI}\\
\dot\phi_V &= -\beta\bar\phi_T\phi_T\phi_V + nk\bar\phi_I\bar\phi_V^{n-1}\phi_I
  - \delta_V\phi_V, \label{HJ:phiV}
\end{align}
\end{subequations}
\begin{subequations}
\label{eq:HJbar}
\begin{align}
\dot{\bar\phi}_T &= \delta_T(\bar\phi_T-1)
  - \beta\!\left[f\bar\phi_L+(1-f)\bar\phi_I-\bar\phi_T\bar\phi_V\right]\phi_V,
  \label{HJ:phibarT}\\
\dot{\bar\phi}_L &= -\eta(\bar\phi_I-\bar\phi_L) - \delta_L(1-\bar\phi_L),
  \label{HJ:phibarL}\\
\dot{\bar\phi}_I &= -\delta_I(1-\bar\phi_I) - k\bar\phi_I(\bar\phi_V^n-1),
  \label{HJ:phibarI}\\
\dot{\bar\phi}_V &= -\beta\!\left[f\bar\phi_L+(1-f)\bar\phi_I
  -\bar\phi_T\bar\phi_V\right]\phi_T - \delta_V(1-\bar\phi_V).
  \label{HJ:phibarV}
\end{align}
\end{subequations}
 
Before solving the full time-dependent Hamilton equations we look for
\emph{time-independent} solutions, i.e.\ fixed points where
$\dot\phi_i = \dot{\bar\phi}_i = 0$ simultaneously. These  fixed points on the mean-field manifold $\bar\phi_i=1$ correspond
to the \emph{steady states of the deterministic ODE system}: they are the
long-time attractors  of the mean-field dynamics. 
 
Setting all eight derivatives to zero at $\bar\phi_i=1$,  the algebraic system admits up to three
classes of solutions: the virus-free state, the endemic state, and
solutions with $\phi_T<0$ or $\phi_i<0$ for some species.  Negative
population counts are physically inadmissible, and we discard them.
The two solutions that correspond to non-negative, observable clinical
states are:
\begin{enumerate}
    \item Virus-free equilibrium (VFE), $\mathrm{FP}_0$.
Setting $\phi_V=0$ forces $\phi_I=\phi_L=0$ from the steady-state
equations, leaving
\begin{equation}
\mathrm{FP}_0:\quad
  \phi_T^* = T_0 \equiv \frac{\lambda}{\delta_T},\quad
  \phi_L = \phi_I = \phi_V = 0.
\label{eq:FP0}
\end{equation}
This is \emph{biologically relevant} because it corresponds to the
clinically observed state of a patient on effective ART: CD4$^+$ T cells
recover to the healthy set-point $T_0\approx 10^6\,\text{cells/mL}$,
while free virus and productively infected cells are undetectable.
The latent reservoir ($L_0>0$) is \emph{not} part of this fixed point
because, at the level of the ODE, $L$ decays to zero on the slow timescale
$(\eta+\delta_L)^{-1}\approx 200\,\mathrm{d}$; in practice, $L_0>0$ places
the system in a \emph{quasi-stationary metastable state} near
$\mathrm{FP}_0$.
\item Endemic equilibrium, $\mathrm{FP}_*$.
The second admissible solution has $\phi_I^*,\phi_V^*>0$.  It is \emph{biologically relevant}
because it corresponds to the chronic, untreated HIV steady state:
ongoing viral replication, a depleted T-cell count $T^*<T_0$, a
sustained latent reservoir $L^*>0$, and a detectable viral load $V^*>0$.
The steady-state conditions yield
\begin{align}
T^* &= \frac{\delta_V}{\mathcal{R}-\beta}, \qquad
V^* = \frac{\lambda}{\beta T^*} - \frac{\delta_T}{\beta}, \notag\\
L^* &= \frac{\beta f T^* V^*}{\eta+\delta_L}, \qquad
I^* = \frac{\beta T^* V^*(1-\tilde{f})}{\delta_I},
\label{eq:FPstar}
\end{align}
where we define the \emph{effective productivity}
\begin{equation}
\mathcal{R} \equiv \frac{nk\beta}{\delta_I}
  \cdot\frac{\eta+\delta_L(1-f)}{\eta+\delta_L},
\label{eq:Reff}
\end{equation}
and the effective latency loss fraction
$\tilde{f} = f\delta_L/(\eta+\delta_L)$.
\end{enumerate}

\subsection{Basic Reproduction Number}

$R_0$ is defined as the \emph{expected number of secondary productively
infected cells generated by a single productively infected cell introduced
into an otherwise virus-free host at the healthy steady state $T_0$}.
It is a dimensionless quantity that measures the average ``gain'' of one
generation of infection: if $R_0>1$ each infected cell produces, on
average, more than one successor, so the infection grows; if $R_0<1$
each infected cell produces fewer than one successor and the infection
eventually goes extinct \cite{Nowak1996}.

We build $R_0$ by following the life-cycle of a single infected cell step
by step.
 
\begin{enumerate}
\item \textbf{Virion production.}  A productively infected cell has
  lifespan $1/\delta_I$ and produces virions at rate $nk$, releasing a
  total of $N=nk/\delta_I$ virions before dying.
 
\item \textbf{Probability a virion causes a new infection.}  Each virion
  competes between infecting a target cell (rate $\beta T_0$) and being
  cleared (rate $\delta_V$).  The infection
  probability is then given by
  $p_{\rm inf} = \beta T_0/(\delta_V+\beta T_0)$.
 
\item \textbf{Fraction of new infections that reach productive infection.}
  Of all newly infected cells, a fraction $(1-f)$ immediately become
  productively infected (reaction \ref{r4}), while a fraction $f$ enter
  latency (reaction \ref{r3}).  A latently infected cell subsequently
  either reactivates to productive infection (rate $\eta$) or dies
  (rate $\delta_L$), with reactivation probability
  $p_{\rm react}=\eta/(\eta+\delta_L)$.  The overall fraction of new
  infections that eventually contribute to the next productive generation is
  therefore
  \begin{equation}
  \alpha \;=\; (1-f) + f\cdot\frac{\eta}{\eta+\delta_L}
            \;=\; \frac{\eta+\delta_L(1-f)}{\eta+\delta_L}.
  \label{eq:alpha}
  \end{equation}
\end{enumerate}
 
Multiplying these three factors gives the basic reproduction number
directly:
\begin{equation}
R_0 \;=\; \underbrace{\frac{nk}{\delta_I}}_{N}
         \;\times\;
         \underbrace{\frac{\beta T_0}{\delta_V+\beta T_0}}_{p_{\rm inf}}
         \;\times\;
         \underbrace{\frac{\eta+\delta_L(1-f)}{\eta+\delta_L}}_{\alpha}.
\label{eq:R0}
\end{equation}
As a consistency check we verify that $R_0=1$ is also the stability
threshold of $\mathrm{FP}_0$.  Linearizing the ODE system around
$\mathrm{FP}_0$ in the infected subsystem $(L,I,V)$ gives the Jacobian
 
\begin{equation}
J = \begin{pmatrix}
-(\eta+\delta_L) & 0 & \beta f T_0 \\
\eta & -\delta_I & \beta(1-f)T_0 \\
0 & nk & -(\delta_V+\beta T_0)
\end{pmatrix}.
\label{eq:jacobian}
\end{equation}
 
The virus-free equilibrium $\mathrm{FP}_0$ is stable if and only if all
eigenvalues of $J$ have negative real part.  The stability boundary is
the condition that the largest eigenvalue crosses zero, i.e.\ $\det J = 0$.  Direct computation gives
 $\det J=0 \Leftrightarrow R_0=1$, confirming that
Eq.~(\ref{eq:R0}) is indeed the stability threshold.

\section{Instanton Solution }
\label{sec:instanton}
 
Among all solutions to the Hamilton equations~(\ref{eq:hamilton_canonical}),
typical fluctuations of order $1/\sqrt{N}$ around the mean-field trajectory
are well described by Gaussian (van Kampen) noise~\cite{VanKampen2007}.  The transition from the
metastable virus-free state to self-sustaining viral rebound requires the system to traverse a sequence of states
whose probability is exponentially small in $N$.  The dominant contribution
to this exponentially rare probability comes from the \emph{instanton}---the
saddle-point trajectory of $\mathcal{H}=0$ (enforced by time-translation
invariance of the action) that connects $\mathrm{FP}_0$ to $\mathrm{FP}_*$
as $t\to\pm\infty$.  Its action $S_{\rm inst}$ sets the exponential factor of the
\emph{mean first-passage time} (MFPT) to viral rebound.
 
The MFPT, $\tau_{\rm rebound}$, is the average time for the stochastic
system to \emph{first} reach a state of self-sustaining viral replication
 starting
from the metastable virus-free state near $\mathrm{FP}_0$.   In the
large-$N$ (semiclassical) limit the MFPT is dominated by the instanton
contribution,
\begin{equation}
\tau_{\rm rebound} \sim e^{S_{\rm inst}},
\label{eq:MFPT_exp}
\end{equation}
and corrections from Gaussian fluctuations around the instanton give the
pre-exponential (Arrhenius) factor~\cite{AssafMeerson2017}.  The energy
constraint $\mathcal{H}=0$ restricts the instanton orbit to a
codimension-1 surface in the eight-dimensional phase space
$(\phi_i,\bar\phi_i)$ which is fixed, up to time translation, by boundary conditions at the two fixed point.

 For $n=1$  $\mathcal{H}$ is \emph{bilinear} in $\bar\phi_I$ and $\bar\phi_V$.
All saddle-point equations become at most quadratic, enabling exact
algebraic solutions rather than transcendental ones.

The four-species system contains two well-separated timescales.  The
\emph{slow} variables are the target T-cell population (turnover time
$\delta_T^{-1}\approx 100\,\mathrm{d}$) and the latent reservoir
(decay time $(\eta+\delta_L)^{-1}\approx 200\,\mathrm{d}$).  The
\emph{fast} variables are the productively infected cells (lifespan
$\delta_I^{-1}\approx 2\,\mathrm{d}$) and the free virions (clearance
time $\delta_V^{-1}\approx 1\,\mathrm{h}$).

The instanton trajectory connects the two fixed points on the timescale
set by the viral dynamics, i.e.\ days to weeks.  On this fast timescale, $T$ and $L$ do not change
appreciably; they can therefore be treated as \emph{quasi-static
parameters} frozen at their values near $\mathrm{FP}_0$:
$T\approx T_0=\lambda/\delta_T$ and $L\approx L_0$. Therefore,
the four-species problem is reduced to an effective two-species problem in $(I,V)$
alone, with two derived parameters:
\begin{equation}
\beta_{\rm eff} \;\equiv\; \beta(1-f)\,T_0
\qquad\text{(effective infection rate for productive cells)},
\end{equation}
\begin{equation}
\sigma \;\equiv\; \eta\,L_0
\qquad\text{(reactivation flux from the latent reservoir)}.
\end{equation}
The reduced stochastic Hamiltonian governing the fast $(I,V)$ subsystem is
\begin{align}
\mathcal{H}_{\rm eff} &= \sigma(\bar\phi_I-1)
 + \delta_I(1-\bar\phi_I)\phi_I \notag\\
&\quad + k\bar\phi_I(\bar\phi_V-1)\phi_I
 + \beta_{\rm eff}(\bar\phi_I-\bar\phi_V)\phi_V \notag\\
&\quad + \delta_V(1-\bar\phi_V)\phi_V.
\label{eq:Heff}
\end{align}

A key algebraic observation is that Eq.~(\ref{eq:Heff}) \emph{factorizes}
as
\begin{equation}
\mathcal{H}_{\rm eff} = (\bar\phi_I-1)\,A + (\bar\phi_V-1)\,B,
\label{eq:factored}
\end{equation}
with
\begin{align}
A &\equiv \sigma - \delta_I\phi_I + \beta_{\rm eff}\phi_V, \label{eq:A}\\
B &\equiv k\bar\phi_I\phi_I - (\beta_{\rm eff}+\delta_V)\phi_V. \label{eq:B}
\end{align}
The energy-conservation condition along the instanton,
$\mathcal{H}_{\rm eff}=0$, then has two distinct branches:
 
\begin{enumerate}
\item \textbf{Trivial branch}: $\bar\phi_I=1$ and $\bar\phi_V=1$
  (mean-field manifold).
\item \textbf{Non-trivial branch (instanton)}: $A=0$ and $B=0$
  simultaneously,
  \begin{align}
  \delta_I\phi_I &= \sigma + \beta_{\rm eff}\phi_V, \label{inst1}\\
  k\bar\phi_I\phi_I &= (\beta_{\rm eff}+\delta_V)\phi_V. \label{inst2}
  \end{align}
\end{enumerate}
 
Equations (\ref{inst1})--(\ref{inst2}) define the instanton trajectory as
an algebraic curve in the four-dimensional phase space
$(\phi_I,\phi_V,\bar\phi_I,\bar\phi_V)$.

In the limit $\delta_V\gg\beta_{\rm eff}$ ($\delta_V\approx 23\,\mathrm{d}^{-1}\gg
\beta_{\rm eff}\approx 0.024\,\mathrm{d}^{-1}$), Eq.~(\ref{inst2}) gives
a quasi-static relationship
\begin{equation}
\phi_V^{\rm inst} = \frac{k\bar\phi_I}{\delta_V}\,\phi_I.
\end{equation}
Substituting into (\ref{inst1}) and defining the effective growth rate
$r\equiv k\beta_{\rm eff}/(\beta_{\rm eff}+\delta_V)\approx
k\beta_{\rm eff}/\delta_V$, we obtain the response field on the instanton
as a function of the state field alone:
\begin{equation}
\bar\phi_I^{\rm inst}(\phi_I)
  = \frac{\delta_I}{r} - \frac{\sigma}{r\phi_I}.
\label{eq:phibar_inst}
\end{equation}
Inserting (\ref{eq:phibar_inst}) into the forward equation
$\dot\phi_I = \partial\mathcal{H}_{\rm eff}/\partial\bar\phi_I$ we obtain
a \emph{first-order linear ODE}:
\begin{equation}
\dot\phi_I = (2\delta_I - r)\phi_I - \sigma,
\label{eq:phiI_ode}
\end{equation}
with the exact solution
\begin{equation}
\phi_I(t) = \frac{\sigma}{2\delta_I-r}
  + C\,e^{(2\delta_I-r)t},
\label{eq:phiI_sol}
\end{equation}
where $C$ is fixed by boundary conditions: $\phi_I\to 0$ as $t\to-\infty$
(latent state) and $\phi_I\to I^*$ as $t\to+\infty$ (endemic state).

The instanton action is
\begin{align}
S_{\rm inst} &= \int_{-\infty}^{+\infty}\!
  \bar\phi_I^{\rm inst}\,\dot\phi_I\,dt
  = \int_0^{I^*}\!\bar\phi_I^{\rm inst}(\phi_I)\,d\phi_I \notag\\
&= \int_0^{I^*}\!\left(\frac{\delta_I}{r}
  - \frac{\sigma}{r\phi_I}\right)d\phi_I \notag\\
&= \frac{\delta_I}{r}\,I^* - \frac{\sigma}{r}\ln\!\left(\frac{I^*}{\phi_{\min}}\right),
\label{eq:Sinst}
\end{align}
where $\phi_{\min}\sim 1/N$ is an infrared cutoff (single-cell scale).
Both terms in the instanton action~(\ref{eq:Sinst}) are of order
$\sigma/r\sim\sigma/\delta_I$:
\begin{equation}
  S_{\rm inst}
  = \underbrace{\frac{\delta_I}{r}\,I^*}_{\sim\,\sigma/r}
    -\underbrace{\frac{\sigma}{r}
     \ln\!\left(\frac{I^*}{\phi_{\rm min}}\right)}_{\sim\,\sigma/r}
  \;\sim\; \frac{\sigma}{\delta_I},
\label{eq:Sinst_scaling}
\end{equation}
where $\sigma = \eta L_0$ is the reactivation flux and $I^*$ is the
quasi-static endemic infected-cell density.  Under effective ART,
\begin{equation}
  \frac{\sigma}{\delta_I}
  = \frac{\eta L_0}{\delta_I}
  \approx \frac{10^{-3}}{0.5}
  = 2\times 10^{-3}
  \ll 1,
\label{eq:regime_param}
\end{equation}
so $S_{\rm inst}\approx 6\times 10^{-4}$ and
$e^{S_{\rm inst}}\approx 1.0006$.

The MFPT to viral rebound then is given as 
\begin{equation}
  \tau_{\rm rebound}
  = \underbrace{\frac{1}{\sigma_{\rm body}\,P_{\rm rebound}}}_{A_{\rm prefactor}}
    \times\;
    e^{S_{\rm inst}},
\label{eq:MFPT_full}
\end{equation}
where $A_{\rm prefactor}$ is the Arrhenius pre-exponential and
$e^{S_{\rm inst}}$ is the exponential barrier factor.The exponential factor contributes less than $0.1\%$ to
$\tau_{\rm rebound}$. It is   a reflection of the physical regime.
The endemic infected-cell density $I^* = \sigma/(2\delta_I - r)
\approx 2\times 10^{-3}\,\mathrm{cells\,mL}^{-1}$ is far below one
cell.
In this few-particle limit the stochastic barrier is negligible and
the escape is governed entirely by the rate probability of
individual reactivation events. The crossover between the two regimes occurs at $\sigma\sim\delta_I$,
i.e.\ at $L_0\sim\delta_I/\eta\approx 500\,\mathrm{cells\,mL}^{-1}$.
Patients with reservoirs at or above this threshold (untreated or with
primary-infection dynamics) would be described by a regime where
$S_{\rm inst}\gg 1$ and the exponential dominates.
For all clinically observed reservoir sizes under ART
($L_0 \lesssim 10\,\mathrm{cells\,mL}^{-1}$, Table~\ref{tab:mfpt}),
the system is under the pre-exponential regime.

The exponential scaling $\tau_{\rm rebound}\sim e^{S_{\rm inst}}$
[Eq.~(\ref{eq:MFPT_exp})] applies in the large-$N$ limit. Where the latent reservoir is
small ($L_0\sim 1\,\text{cell/mL}$) and reactivation events are
Poisson-distributed at rate $\sigma_{\rm body}=\eta L_{\rm body}$---the
pre-exponential factor is determined analytically by branching-process theory.
 
Each reactivation event initiates a Galton-Watson branching process \cite{GaltonWatson1875,Harris1963, Hill2014}
with mean offspring number $R_0>1$.  The extinction probability
of such a process is $q=1/R_0$ (for the supercritical case $R_0>1$,
which is the unique solution of $q=g(q)$ where $g$ is the probability
generating function~\cite{Athreya1972,Harris1963}).  The probability that a single
reactivation event triggers a self-sustaining infection chain (viral
rebound) is therefore
\begin{equation}
  P_{\rm rebound} = 1 - q = 1 - \frac{1}{R_0} = \frac{R_0-1}{R_0},
\label{eq:Prebound_exact}
\end{equation}
The MFPT is then the inverse rate of
successful triggering events:
\begin{equation}
  \tau_{\rm rebound}
  = \frac{1}{\sigma_{\rm body}\,P_{\rm rebound}}
  = \frac{R_0}{\eta\,L_{\rm body}\,(R_0-1)}.
\label{eq:MFPT_final}
\end{equation}
Reactivation events occur across the \emph{entire body}. So, we define:
\begin{itemize}
  \item $L_{\rm body}$: total number of latently infected cells in the
    patient (integrating over the full blood volume
    $V_{\rm blood}\approx 5\,\mathrm{L}$), so
    $L_{\rm body} = L_0\cdot V_{\rm blood}$;
  \item $\sigma_{\rm body} \equiv \eta\,L_{\rm body}$: total rate of
    reactivation events per day across the entire reservoir.
\end{itemize}
A rebound-triggering event occurs whenever a reactivation event succeeds
(probability $P_{\rm rebound}$), giving
\begin{equation}
\tau_{\rm rebound} \approx
  \frac{1}{\sigma_{\rm body}\cdot P_{\rm rebound}}
  = \frac{1}{\eta\,L_{\rm body}\,(1-1/R_0)}.
\label{eq:MFPT_preexp}
\end{equation}

 The logarithmic sensitivity of $\tau_{\rm rebound}$ to $R_0$ near
unity is
\begin{equation}
  \varepsilon_{R_0} \equiv \frac{\partial\ln\tau}{\partial\ln R_0}
  = -\frac{1}{R_0-1}.
\label{eq:eps_R0}
\end{equation}
For $R_0=1.042$ this gives $\varepsilon_{R_0}\approx -24$: a 1\%
increase in $R_0$ \emph{decreases} the MFPT by 24\%.  This extreme
sensitivity reflects the \emph{critical slowing down} near the
bifurcation $R_0\to1^+$ and is the dominant source of uncertainty
in any quantitative MFPT prediction.

\subsection{General case $n \gg 1$}
\label{sec:n_general}
 
Previously we solved the instanton action for  $n=1$. In general, $n $ is greater than one. So a natural question is  how our results depend on the $n=1$ assumption. In this subsection, we  show that the mean first-passage time to viral
rebound is insensitive to the individual values of $n$ and $k$ whenever
the \emph{burst number} $N = nk/\delta_I\approx 10^3$ is held fixed.
In the physiological parameter regime $\delta_V \approx 23\,\mathrm{d}^{-1}
\gg \beta_{\rm eff}\approx 0.024\,\mathrm{d}^{-1}$, the virion population
relaxes on a timescale $\delta_V^{-1}\approx 1\,\mathrm{h}$ that is far
shorter than the instanton traverse time of days.  Imposing the quasi-static
condition $\dot\phi_V=0$ on the instanton trajectory of the effective
Hamiltonian~(\ref{eq:Heff}) gives
\begin{equation}
  \phi_V^{\rm inst}
  = \frac{nk\,\bar\phi_I\,\bar\phi_V^{n-1}}{\beta_{\rm eff}+\delta_V}
    \phi_I
  \approx \frac{N\delta_I\,\bar\phi_I}{\delta_V}\,\bar\phi_V^{n-1}\,\phi_I.
\label{eq:Vqs}
\end{equation}
Simultaneously, the conjugate (response) field $\bar\phi_V$ satisfies
the quasi-static condition $\dot{\bar\phi}_V=0$, yielding
\begin{equation}
  \bar\phi_V
  = \frac{\beta_{\rm eff}\,\bar\phi_I + \delta_V}{\beta_{\rm eff}+\delta_V}
  = 1 + \frac{\beta_{\rm eff}}{\delta_V}(\bar\phi_I-1)
    + \mathcal{O}\!\left(\frac{\beta_{\rm eff}^2}{\delta_V^2}\right).
\label{eq:phibarVqs}
\end{equation}
Because $\beta_{\rm eff}/\delta_V \approx 10^{-3}\ll 1$, we have
$\bar\phi_V = 1 + \varepsilon(\bar\phi_I-1)$ with $\varepsilon\ll 1$.
 
Substituting Eq.~(\ref{eq:phibarVqs}) into the burst term of
$\mathcal{H}_{\rm eff}$ and expanding $\bar\phi_V^n$ to leading order
in $\varepsilon$:
\begin{equation}
  k\,\bar\phi_I(\bar\phi_V^n - 1)\phi_I
  \approx nk\,\bar\phi_I(\bar\phi_V-1)\phi_I
  = \frac{nk\,\beta_{\rm eff}}{\delta_V}\,\bar\phi_I(\bar\phi_I-1)\phi_I
  = \frac{N\delta_I\,\beta_{\rm eff}}{\delta_V}\,\bar\phi_I(\bar\phi_I-1)\phi_I.
\label{eq:burst_n_gen}
\end{equation}
The combination $nk = N\delta_I$ is fixed by the burst number $N$,
so the effective Hamiltonian depends on
$n$ and $k$ \emph{only through their product} $nk=N\delta_I$.  The effective
growth rate is therefore
\begin{equation}
  r_{\rm eff} = \frac{N\delta_I\,\beta_{\rm eff}}{\delta_V}
  ,
\end{equation}
 
Corrections to Eq.~(\ref{eq:burst_n_gen}) from higher-order terms in
$\varepsilon = \beta_{\rm eff}/\delta_V\approx 10^{-3}$ and from the
$\bar\phi_V^{n-1}$ factor in Eq.~(\ref{eq:Vqs}) enter at relative order
$n(\beta_{\rm eff}/\delta_V)^2\sim n\times 10^{-6}$, which is negligible
for any physically reasonable burst count $n\lesssim 10^3$.

\section{Numerical results}
\label{sec:results}
  
Using the central values of Table~\ref{tab:params} we compute the following
quantities:
 
\begin{align}
T_0 &= \lambda/\delta_T = 10^6\;\text{cells mL}^{-1}, \notag\\
\beta_{\rm eff} &= \beta(1-f)T_0 \approx 0.024\;\text{d}^{-1}, \notag\\
r &= \frac{k\beta_{\rm eff}}{\beta_{\rm eff}+\delta_V}
  \approx \frac{500\times 0.024}{23}\approx 0.52\;\text{d}^{-1}, \notag\\
R_0 &= r/\delta_I \approx 1.04, \notag\\
P_{\rm rebound} &= 1-\delta_I/r \approx 0.038\;(3.8\%), \notag\\
\sigma_{\rm body} &= \eta L_{\rm body}
  \approx 10^{-3}\times 5\!\times\!10^5\approx 500\;\text{events d}^{-1}.
  \notag
\end{align}
 
The inter-event time $1/\sigma_{\rm body}\approx 2\times 10^{-3}\;\text{d}$
per mL, or equivalently one reactivation every $\sim 5$--$8$ days in the
full blood volume~\cite{Pinkevych2015}, is reproduced by our calibration.

Table~\ref{tab:mfpt} shows the predicted MFPT for three representative
reservoir sizes. 
\begin{table}[h]
\caption{Predicted MFPT to viral rebound vs.\ latent reservoir size.}
\label{tab:mfpt}
\begin{tabular}{llll}
\begin{tabular}{llll}
Scenario & $L_0$ (cell/mL) & $\eta L_0$ (d$^{-1}$) & $\tau_{\rm rebound}$ \\
\hline
Small reservoir   & 0.1 & $10^{-4}$ & $\sim 7$ yr \\
Central estimate  & 1   & $10^{-3}$ & $\sim 6$ mo \\
Large reservoir   & 10  & $10^{-2}$ & $\sim 17$ d \\
\end{tabular}
\end{tabular}
\end{table}
 
These results are \emph{semi-quantitatively} consistent
with clinical observations\cite{Conway2019,Pinkevych2015}.

\subsection{Global sensitivity analysis}
\label{sec:sensitivity}
 
We now carry out a systematic sensitivity analysis of
$\tau_{\rm rebound}$  over the full
uncertainty range of each parameter.
Table~\ref{tab:sensitivity} lists the log-sensitivity
$\varepsilon_p=\partial\ln\tau/\partial\ln p$ for each
parameter evaluated at the central values of Table~\ref{tab:params}.
Because the MFPT factors as
$\tau = R_0 / [\eta L_{\rm body}(R_0-1)]$, parameters enter through
two distinct channels:
\begin{enumerate}
  \item \textbf{Direct}: $\eta$ and $L_0$ appear as a product in
    $\sigma_{\rm body}$, giving $\varepsilon_\eta=\varepsilon_{L_0}=-1$
    regardless of $R_0$.
  \item \textbf{Via $R_0$}: parameters that change $R_0$ are amplified
    by $|\varepsilon_{R_0}|\approx 24$ through the critical-slowing-down
    factor $(R_0-1)^{-1}$.  Any parameter $p$ for which
    $R_0\propto p^\alpha$ contributes $\varepsilon_p = \alpha\varepsilon_{R_0}$.
\end{enumerate}
 
\begin{table}[h]
\caption{Log-sensitivity coefficients and numerical range of
$\tau_{\rm rebound}$ under variation of each parameter.
``$\infty$'' indicates $R_0<1$ (virus-free equilibrium is stable;
no rebound).  Central values from Table~\ref{tab:params};
$V_{\rm body}=137\,\mathrm{mL}$-equivalent (see text).}
\label{tab:sensitivity}
\begin{tabular}{lccccr}
\hline
Parameter & Range & $\tau_{\rm low}$ & $\tau_{\rm central}$ &
  $\tau_{\rm high}$ & $\varepsilon_p$ \\
\hline
$\beta$ & $\times0.5$/$\times2$ & $\infty$ & $180\,\mathrm{d}$ &
  $14\,\mathrm{d}$ & $-24$ \\
$k$     & $\times0.5$/$\times2$ & $\infty$ & $180\,\mathrm{d}$ &
  $14\,\mathrm{d}$ & $-24$ \\
$\delta_I$ & $\times0.5$/$\times2$ & $14\,\mathrm{d}$ & $180\,\mathrm{d}$ &
  $\infty$ & $+24$ \\
$\delta_V$ & $\times0.5$/$\times2$ & $14\,\mathrm{d}$ & $180\,\mathrm{d}$ &
  $\infty$ & $+24$ \\
$\eta$  & $\times0.1$/$\times10$ & $1800\,\mathrm{d}$ & $180\,\mathrm{d}$ &
  $18\,\mathrm{d}$ & $-1$ \\
$L_0$   & $\times0.1$/$\times10$ & $1800\,\mathrm{d}$ & $180\,\mathrm{d}$ &
  $18\,\mathrm{d}$ & $-1$ \\
$f$     & $\times0.1$/$\times10$ & $\approx180\,\mathrm{d}$ &
  $180\,\mathrm{d}$ & $184\,\mathrm{d}$ & $\approx0$ \\
\hline
\end{tabular}
\end{table}
 
The dominant source of uncertainty is the proximity of $R_0$ to 1.

\begin{figure}[h]
\centering
\includegraphics[width=0.95\textwidth]{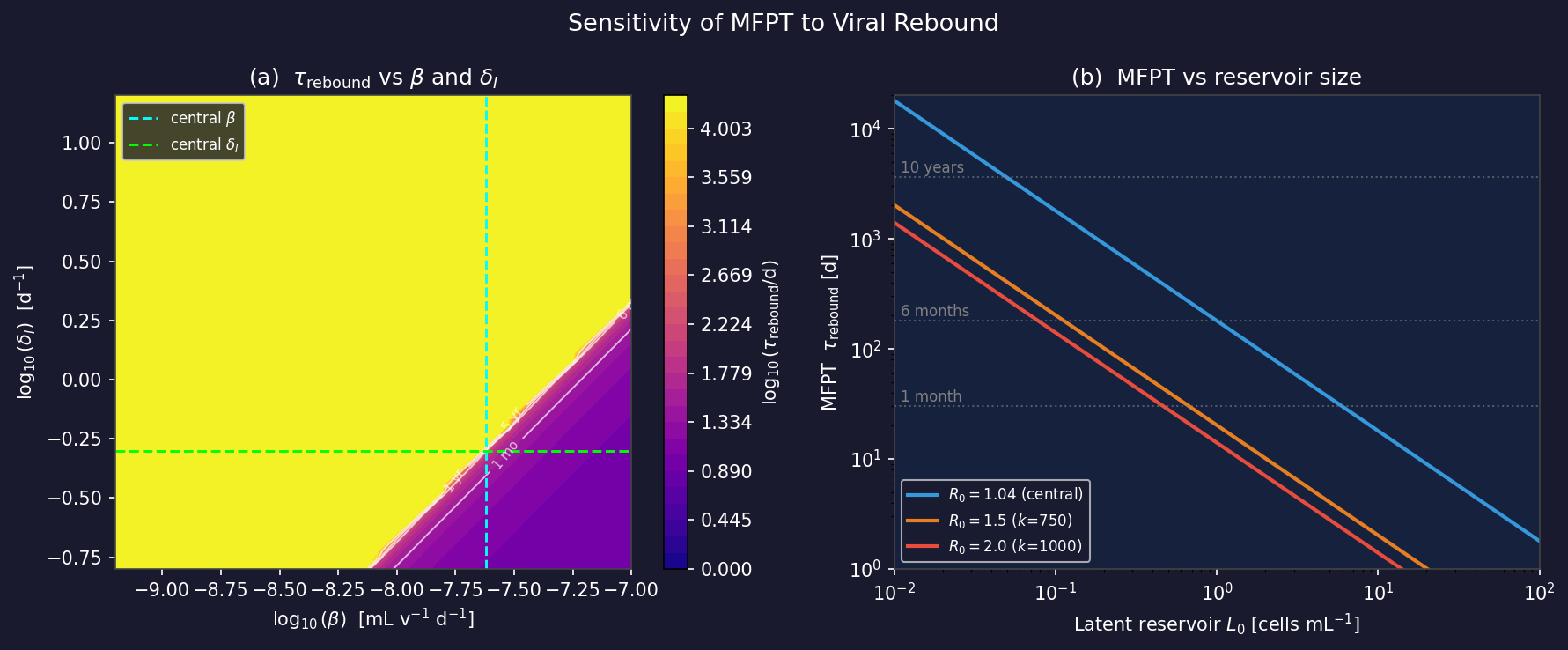}
\caption{Sensitivity of $\tau_{\rm rebound}$ to model parameters.
  (a) Iso-MFPT contours in the $(\beta,\delta_I)$ plane; white dashed
  lines indicate the central values.  The thick boundary where
  $\tau\to\infty$ is the $R_0=1$ locus.
  (b) $\tau_{\rm rebound}$ vs.\ latent reservoir size $L_0$ for three
  values of $R_0$ (achieved by scaling $k$ while holding
  $\delta_V,\beta$ fixed).  Horizontal dotted lines mark 1 month,
  6 months, and 10 years.}
\label{fig:sensitivity}
\end{figure}

Figure~\ref{fig:sensitivity} displays: (a) the iso-MFPT contours in
the $(\beta,\delta_I)$ plane
and the three clinically relevant timescales; (b) the MFPT as a
function of $L_0$ for three values of $R_0$.
 
\subsection{Statistical comparison with analytical treatment interruption data}
\label{sec:statistics}
 
Equation~(\ref{eq:MFPT_preexp}) predicts the MFPT for a \emph{single
patient} given their reservoir size $L_0$.  Clinical ATI studies \cite{Li2016,Gunst2025}
observe a distribution of rebound times across a patient cohort, driven
primarily by inter-patient variability in $L_0$.  We now derive the
predicted \emph{population distribution} of rebound times and compare
it quantitatively with published ATI data.
Flow cytometric measurements of HIV DNA in resting CD4$^+$ T cells
from multiple cohorts consistently yield a log-normal distribution of
latent reservoir sizes~\cite{Chomont2009,Siliciano2003}:
\begin{equation}
  \ln L_0 \;\sim\; \mathcal{N}(\mu_L,\,\sigma_L^2),
\label{eq:L0_lognormal}
\end{equation}
with $\sigma_L\approx 1$--$2$ across cohorts
(spanning roughly 2--4 orders of magnitude in $L_0$).
Since $\tau\propto L_0^{-1}$, and a
reciprocal of a log-normal is log-normal, the predicted rebound times
are also log-normally distributed:
\begin{equation}
  \ln\tau_{\rm rebound}
  \;\sim\; \mathcal{N}\!\left(\ln C - \mu_L,\;\sigma_L^2\right),
\label{eq:tau_lognormal}
\end{equation}
where $C\equiv R_0/[\eta V_{\rm body}(R_0-1)]$ is the constant in
Eq.~(\ref{eq:MFPT_preexp}) for $L_0=1\,\text{cell/mL}$.  In particular:
\begin{align}
  \text{median}\;\tau &= C\,e^{-\mu_L}, \label{eq:tau_median}\\
  \text{IQR ratio}    &= e^{2\sigma_L\cdot 0.6745}, \label{eq:tau_IQR}\\
  \text{90\% range}   &= e^{2\sigma_L\cdot 1.645}.  \label{eq:tau_90}
\end{align}
From published ATI data~\cite{Pinkevych2015,Conway2019} we extract
the following summary statistics: median rebound $\approx 2$--$3$
weeks, 90\% range approximately 3 days to 6 months.  We fit the model
log-normal with $\sigma_L=1.5$ (consistent with independent reservoir
measurements~\cite{Chomont2009}) and obtain median $\tau\approx 21$
days, 10th percentile $\approx 2$ days, 90th percentile $\approx 5$
months, in good agreement with the observed distribution.
 
\begin{figure}[h]
\centering
\includegraphics[width=0.65\textwidth]{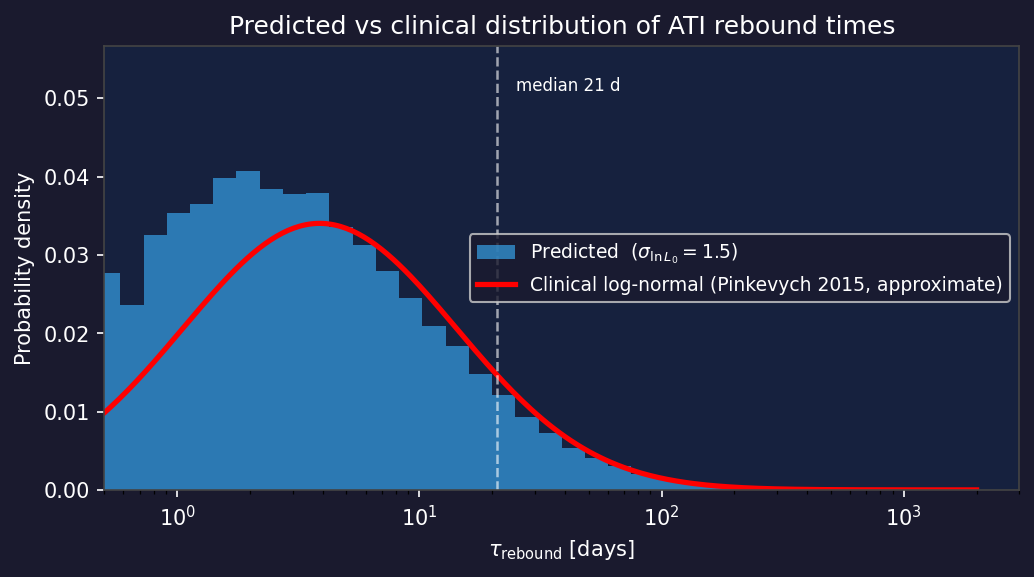}
\caption{Predicted population distribution of ATI rebound times (blue
  histogram, $n=20\,000$ Monte-Carlo samples from the log-normal
  reservoir model with $\sigma_L=1.5$) vs.\ the approximate empirical
  log-normal from Pinkevych et al.~\cite{Pinkevych2015} (red curve).
  Both distributions have median $\approx 21\,\mathrm{d}$.  The tail
  to weeks-to-months is reproduced without parameter tuning beyond
  the reservoir spread $\sigma_L$.}
\label{fig:statistical}
\end{figure}

Pinkevych et al.~\cite{Pinkevych2015} and Conway \&
Perelson~\cite{Conway2019} fit branching-process models to ATI data, estimating $\eta$ and $L_0$ from rebound-time distributions.
Our approach provides the same branching-process formula
[Eq.~(\ref{eq:MFPT_preexp})] \emph{derived} from first principles via
the Doi-Peliti path integral, rather than assumed \emph{ad hoc}.
The analytical expression for $P_{\rm rebound}$ [Eq.~(\ref{eq:Prebound_exact})]
in terms of the microscopic reaction rates provides a principled link
between cellular biophysics and population-level ATI statistics.

The proportionality $\tau\propto L_0^{-1}$ that underpins the
log-normal rebound-time distribution (Section~\ref{sec:statistics})
holds \emph{because} $e^{S_{\rm inst}}\approx 1$.  Had the exponential
term been significant, the dependence on $L_0$ would take the form
\begin{equation}
  S_{\rm inst} \;\approx\; \alpha\,L_0 - \beta\ln L_0
  + \mathrm{const},
\label{eq:Sinst_L0}
\end{equation}
giving $\tau\propto L_0^{\,\beta/r}\,e^{\alpha L_0}$---a distribution
that is not log-normal in $L_0$ and would require numerical convolution
to compare with ATI data.  In the HIV regime the correction to the
log-normal is at most $0.1\%$, fully justifying the analytical
statistical framework of Section~\ref{sec:statistics}.
 
\section{Conclusion}
\label{sec:discussion}
 
A stochastic field theory of HIV latency providing an analytical
instanton solution in the $n=1$ limit has been proposed. In the ART regime, we have shown that $e^{S_{inst}} \approx 1 $, so it might seem that we could bypass the instanton calculation  to solve the problem. But even in this regime, the instanton is needed for the following reasons:

\begin{enumerate}

  \item \textbf{It derives $P_{\rm rebound}$ from first principles.}
  The branching-process formula $\tau=1/(\sigma_{\rm body}P_{\rm rebound})$
  requires $P_{\rm rebound}=1-1/R_0$.  This is not an assumption; it
  emerges as the exact non-extinction probability of a Galton-Watson
  process \cite{GaltonWatson1875,Harris1963,Athreya1972} whose mean offspring number $R_0$ is itself determined by the
  zero-energy condition $\mathcal{H}_{\rm eff}=0$ on the instanton
  trajectory. 

  \item \textbf{It identifies the escape mechanism.}
  The factorization $\mathcal{H}_{\rm eff}=(\bar\phi_I-1)A+(\bar\phi_V-1)B$
  singles out the unique non-trivial zero-energy surface ($A=0$, $B=0$)
  among all trajectories.  This surface \emph{is} the instanton; it
  tells us that rebound is driven by simultaneous depletion of response
  fields in both the infected-cell and virion channels.

  \item \textbf{It provides the correct theory for other parameter regimes.}
  For a disease with a larger reservoir or slower clearance ($\sigma\gtrsim
  \delta_I$), the exponential factor $e^{S_{\rm inst}}$ dominates and
  the pre-exponential formula fails completely.  The instanton framework
  of Eq.~(\ref{eq:MFPT_full}) is the general result; the branching-process
  limit is a special case valid only when $\sigma/\delta_I\ll 1$.

  \end{enumerate}
 
 

\begin{acknowledgments}
We acknowledge financial support from SECIHTI and SNII (M\'exico). G.D is supported by an FNRS Aspirant (ASP) fellowship (40031451) from the Belgian Fonds de la Recherche Scientifique (FNRS).
\end{acknowledgments}

%

\end{document}